\documentclass{article}
\title{Elementary Theory of Line Broadening and Four-wave Mixing in
Nonequilibrium Many-Particle Systems}
\author{David W. Snoke\thanks{email: snoke@pitt.edu}\\
Department of Physics and Astronomy, University of Pittsburgh\\
3941 O'Hara St.\\
Pittsburgh, PA 15260\\
}
\begin{document}

\maketitle

\begin{center}
\parbox{12cm}{{\bf Abstract.} The basic results of optical line broadening and
four-wave mixing are deduced from first principles based on 
time-dependent many-body
perturbation theory. The formalism allows us to write all the results 
in terms of
nonequilibrium distribution functions without the need to assume a 
quasiequilibrium
temperature. The connection of these results to the quantum Boltzmann 
equation is
shown, which is also derived from first principles. Last, specific 
predictions for
electron-electron scattering are reviewed.}
\end{center}
\vspace{1cm}

\newpage
There are many optical experiments which measure linewidths, lifetimes, $T_1$
times and $T_2$ times, etc., but it is not always clear how these relate to
each other in terms of the basic quantum mechanics. This paper gives a basic
introduction to these experiments based on elementary time-dependent many-body
perturbation theory in the random-phase approximation. Of primary 
interest is to be
able to write down this theory for the general case of a 
nonequilibrium many-body
system without assuming an equilibrium temperature. We do not want 
even to assume a
perturbation around a quasiequilibrium system, but a system which is far from
equilibrium, e.g. a shell in momentum space. This type of highly nonequilibrium
system is often created in ultrafast laser experiments.

It turns out that it is nontrivial to show a very basic experimental 
result, that
optical lines are broadened in the shape of a Lorentzian which has width
proportional to the instantaneous scattering rate, in the case of a 
nonequilibrium
many-body system. The assumptions necessary to deduce this result in 
turn show the
limits of its applicability.

\section{Time-Dependent Perturbation Theory}
\label{sect1}

The theory here is based on standard time-dependendent
perturbation theory, e.g. as presented by Baym \cite{baym}. We review here the
basic results given in Chapter 12 of Baym.

We begin by assuming that the Hamiltonian of a quantum mechanical system is
given by
$$
H = H_0 + V,
$$
where $H_0$ is the single-particle Hamiltonian and $V$ is a scattering term
which is small compared to $H_0$.  The system of interest can be anything from
free atoms or ions to carriers in a semiconductor or metal. We ignore all
relativistic effects, however.

We assume that the system has been prepared at time $t=0$ in
the quantum mechanical state  $| \psi_0 \rangle$.  At a later time $t$, the
state is written as $| \psi_t \rangle$.

The Schr\"odinger equation gives the time evolution of the system as
$$
i\hbar \frac{\partial}{\partial t}| \psi_t \rangle = (H_0 + V)| \psi_t \rangle
$$
In the interaction representation, we define a new state $| \psi(t) \rangle$
($t$ in parenthesis rather than subscript) given by
$$
| \psi(t) \rangle = e^{iH_0t/\hbar}| \psi_t \rangle
$$
and a new operator $V(t) = e^{iH_0t/\hbar}Ve^{-iH_0t/\hbar}$. In this
representation, the Schr\"odinger equation is rewritten as
$$
i\hbar \frac{\partial}{\partial t}| \psi(t) \rangle =  V(t)| \psi(t) \rangle,
$$
which has the advantage of not depending on $H_0$.

Integrating this equation as a perturbation series, one can show that
\cite{baym}
\begin{eqnarray}
| \psi(t) \rangle &=& e^{-(i/\hbar)\int V(t) dt} | \psi(0)
\rangle\label{t1} \\ & = & \left( 1 + (1/i\hbar) \int_{0}^t V(t')dt'
+   (1/i\hbar)^2 \int_{0}^t dt' \int_{0}^{t'} dt'' V(t')V(t'')+ ...
\right)| \psi(0) \rangle
\nonumber
\end{eqnarray}

\section{Connection to Time-Independent Perturbation Theory}
\label{sect2}

In general, we are interested in knowing the amount of depletion of the
initial state after some time $t$. Abbreviating
$|\psi(0) \rangle = | 0 \rangle$, we write
\begin{eqnarray}
\langle 0| \psi(t) \rangle & = & \left( 1 + (1/i\hbar) \langle 0 | V | 0
\rangle \int_{0}^t  dt'    \right. \nonumber \\
&&+ (1/i\hbar)^2 \int_{0}^t dt' \int_{0}^{t'} dt''
\sum_m \langle 0 | V| m \rangle \langle m | V | 0 \rangle
e^{(i/\hbar)(E_0-E_m)t'}e^{(i/\hbar)(E_m-E_0)t''}
\nonumber \\ &&  +(1/i\hbar)^3 \int_{0}^t dt'
\int_{0}^{t'} dt''
\int_{0}^{t''} dt'''
\sum_{m,n} \langle 0 | V| n \rangle \langle n | V| m \rangle \langle m | V| 0
\rangle \nonumber \\ &&
\times
e^{(i/\hbar)(E_0-E_n)t'}e^{(i/\hbar)(E_n-E_m)t''}e^{(i/\hbar)(E_m-E_0)t'''}
  + ... \biggr) \label{e1}.
\end{eqnarray}
Here we have inserted a sum over the complete set of states $\sum | m \rangle
\langle m| = 1$ which are assumed to be eigenstates of the Hamiltonian $H_0$.
Then we have
\begin{eqnarray}
\langle 0| \psi(t) \rangle &=& \biggl(1 +  (1/i\hbar)t \langle 0 | V | 0
\rangle  + \frac{1}{2}(1/i\hbar)^2t^2 |\langle 0 | V | 0 \rangle|^2 \nonumber
\\ &&  + (1/i\hbar)   \sum_{m\ne 0} \int_{0}^t dt' \frac{|\langle m | V | 0
\rangle|^2}{E_0-E_m+i\eta}
e^{(i/\hbar)(E_0-E_m)t'}\left( e^{(i/\hbar)(E_m-E_0)t'}-1 \right)
\nonumber \\ &&
+ \frac{1}{3!}(1/i\hbar)^3t^3 |\langle 0 | V | 0 \rangle|^3
\nonumber \\ &&
+ 2(1/i\hbar)^2\langle 0 | V | 0 \rangle \sum_{m\ne 0}
  \int_{0}^t dt'\int_{0}^{t'} dt'' \frac{|\langle m | V | 0
\rangle|^2}{E_0-E_m+i\eta}
e^{(i/\hbar)(E_0-E_m)t'}\nonumber \\ &&
\times\left( e^{(i/\hbar)(E_m-E_0)t'}-1 \right)
+ ... \biggr) \label{e2}.
\end{eqnarray}
In going from (\ref{e1}) to (\ref{e2}), we have used some assumptions.
The upper and lower bounds of each time integral give terms of the
form
$$
\left( e^{(i/\hbar)(E_m-E_0)t'}-1 \right).
$$
The first, exponential term cancels the  exponential term of the next time
integral. The second term, -1, gives a vanishing term in the
integration because when the last time integration over $dt'$
is performed, it gives an integral of the form
$$
\int {\cal D}(E_m) dE_m \left(e^{i(E_m-E_0)t/\hbar} -1\right) 
\frac{|\langle m |
V | 0 \rangle|^2}{(E_0-E_m)^2}.
$$
Assuming ${\cal D}(E_m)|\langle m | V | 0 \rangle|^2$ is continuous and slowly
varying with $E_m$, this integral vanishes because the leading term 
is odd, i.e.
$$
\left(e^{i(E_m-E_0)t/\hbar} -1\right) \simeq i(E_m-E_0)t/\hbar
$$
for $(E_m-E_0) \simeq 0$, and when $(E_m-E_0) \gg 0$, the $1/(E_0-E_m)^2$ term
in the denominator and the fast oscillation of the $e^{i(E_m-E_0)t/\hbar}$ term
kill this integral. This is the ``random-phase approximation'' (RPA);
essentially, it means that we ignore memory of the past and keep only terms
from the upper bound of the time integrals.

Note also that rigorously, to introduce the term
$i\eta$ in the denominator, which allows us to treat the pole at $E_m=E_0$, we
must assume that $V=Ve^{-\eta t}$, where
$\eta
\rightarrow 0$. Then we will have
$$
\int_{0}^t e^{-\eta t'}dt' = \frac{e^{-\eta t}-1}{-\eta} = \frac{1-\eta
t + ... - 1}{-\eta} = t
$$
and
\begin{eqnarray}
\int_{0}^t \int_{0}^{t'} e^{-\eta t'} e^{-\eta t''}dt' dt'' =
\int_{0}^t e^{-\eta t'}\frac{e^{-\eta t'}-1}{-\eta} dt' =
\frac{e^{-2\eta t}-1}{2\eta^2}-\frac{e^{-\eta t}-1}{\eta^2} \nonumber\\
=
\frac{(1-2\eta t+2\eta t^2+ ... - 1) - 2(1 - \eta t + \frac{1}{2}t^2 +... -1)
}{2\eta^2} = \frac{1}{2} t^2, \nonumber
\end{eqnarray}
etc., which gives us the same result as if we had simply done the integrals
assuming $V$ is time-independent and then inserted $i\eta$ whereever it is
needed to take care of a pole.

The series in (\ref{e2}) has the same form as the series expansion of an
exponential. By induction, we can write
\begin{equation}
\langle 0| \psi(t) \rangle = \exp\left[-(i/\hbar)\left( \langle 0 | V |
0 \rangle +  \sum_{m\ne 0}
  \frac{|\langle m | V | 0 \rangle|^2}{E_0-E_m+i\eta} + ... \right) t
\right]
\label{exp}
\end{equation}
This result (\ref{exp}) is extremely useful. It is not an approximation,
but is exact to all orders of $t$, within the limits of the RPA, for any
time-independent $V$. The series inside the exponential is just the
time-independent perturbation series for the energy correction due to the
particle interactions (see, e.g. Baym \cite{baym}, chapter 11), which is
typically called ``Rayleigh-Schr\"odinger'' perturbation theory.

By the Dirac formula, the second-order term is
\begin{eqnarray}
\sum_{m\ne 0}
  \frac{|\langle m | V | 0 \rangle|^2}{E_0-E_m+i\eta}
&=& P\left( \sum_{m\ne 0}\frac{|\langle m | V | 0 \rangle|^2}{E_0-E_m}\right)
-  i\pi\sum_{m\ne 0}|\langle m | V | 0
\rangle|^2 \delta(E_0-E_m)\nonumber\\ &=& \Delta^{(2)} - i\Gamma^{(2)}
\label{dirac}
\end{eqnarray}
where the second term is just the total scattering rate.
$\Delta^{(1)} = \langle 0 | V | 0 \rangle$ is called the mean-field energy,
$\Delta^{(2)}$ is called the real self-energy and
$\Gamma^{(2)}$ is called the imaginary self-energy. From (\ref{exp}) we have
therefore
\begin{equation}
\langle 0| \psi(t) \rangle = e^{-(i/\hbar)(\Delta^{(1)}+\Delta^{(2)} -
i\Gamma^{(2)})t}
\label{expg}
\end{equation}
The probability of being in the state $| 0 \rangle$, given by $|\langle 0 |
\psi(t)
\rangle|^2$, decreases over time as $e^{-(2\Gamma/\hbar)t}$ due to
out-scattering. Thus one can see the reason why the imaginary self-energy is
associated with an out-scattering rate.

\section{Absorption and Emission Line Shape in the Single-Particle Picture}
\label{sect3}

We now imagine that we have a single particle prepared in a given state which
can decay by two channels, namely by coupling to an external field (normally a
photon field) and by scattering to other internal states. We write
$$
H = H_0 + V + V'
$$
where $V'$ is a smaller perturbation than $V$, which gives the
interband photon-electron interaction. We will only be concerned about the
first-order contributions of this term, i.e. we will not worry about
renormalization of the electron states due to electron-photon interaction. We
will allow renormalization due to $V$, however. Therefore we write
$$
|\psi'(t)\rangle = |\psi(0)\rangle
+ \frac{1}{i\hbar}\int_0^t dt' V'(t')|\psi(t')\rangle,
$$
where $|\psi(t')\rangle$ is assumed to include the evolution due to $V$ given
in (\ref{t1}).

We are interested in the rate of emission of a photon with frequency
$\omega$ and momentum ${\bf k}$. We assume that the eigenstates of $H_0$
are single-particle states, each with a given momentum. Therefore by
momentum conservation, only one state with momentum ${\bf k}$, which we will
identify as $| 0
\rangle$, couples to the photon state
$|\omega,{\bf k}\rangle$. We write the rate of emission as the probability
of being in the photon state $|\omega,{\bf k}\rangle $ at $t=\infty$, after
starting in the initial state $|0\rangle$, divided by the total time spent in
the initial state:
$$
\frac{1}{\tau(\omega,{\bf k})} =  \frac{\left| \langle  \omega,{\bf
k} | \psi'(\infty)  \rangle  \right|^2}{\int_0^{\infty}dt |\langle 0 | \psi(t)
\rangle |^2}.
$$
By the assumptions above, by momentum conservation we write $\langle
\omega,{\bf k} | V' | m \rangle = \langle V' \rangle \delta_{0,m}$.
  Then we have
\begin{eqnarray}
\left| \langle  \omega,{\bf
k} | \psi'(\infty)  \rangle  \right|^2 &=&
\left|\frac{1}{i\hbar}\int_0^{\infty} dt' \sum_m \langle
\omega,{\bf k} | e^{(i/\hbar)H_0 t'} V' e^{-(i/\hbar)H_0 t'}|  m \rangle
\langle m|  \psi(t')
\rangle
\right|^2\nonumber \\
&=&  \left|\frac{1}{i\hbar} \int_0^{\infty} dt'\langle
\omega,{\bf k} |  V' |  0 \rangle e^{i\omega t'}e^{-(i/\hbar)E_0 t'} \langle 0|
\psi(t')
\rangle
\right|^2\nonumber \\
&=& |\langle V' \rangle|^2 \left|\frac{1}{i\hbar}\int_0^{\infty} dt  \
e^{i\omega t'} e^{-(i/\hbar)E_0 t'} e^{-(i/\hbar)(\Delta^{(1)}+\Delta^{(2)} -
i\Gamma^{(2)})t'}\right|^2\nonumber\\ &=&|\langle V' \rangle|^2 \left|
\frac{1}{\hbar\omega - E -i\Gamma}
\right|^2\nonumber\\
&=& |\langle V' \rangle|^2 \frac{1}{(\hbar\omega - E)^2+\Gamma^2}\nonumber
\end{eqnarray}
where we have abbreviated $E = E_0 + \Delta^{(1)}+\Delta^{(2)}$ and $\Gamma =
\Gamma^{(2)}$ from (\ref{expg}). The normalization factor is
\begin{eqnarray}
\int_0^{\infty}dt \left|\langle 0 | \psi(t)  \rangle \right|^2 & =&
\int_0^{\infty}dt \left|e^{-(i/\hbar)(\Delta^{(1)}+\Delta^{(2)} - 
i\Gamma^{(2)})
t}\right|^2
\nonumber\\
&=& \int_0^{\infty}dt e^{-2\Gamma t/\hbar} \nonumber\\
&=& \frac{\hbar}{2\Gamma} \nonumber
\end{eqnarray}
This gives the total rate as
\begin{equation}
\frac{1}{\tau(\omega,{\bf k})} = |\langle V' \rangle|^2
\frac{2\Gamma/\hbar}{(\hbar\omega - E)^2+\Gamma^2}
\label{rate}
\end{equation}
which when $\Gamma \rightarrow 0$, is
$$
\frac{1}{\tau(\omega,{\bf k})} = \frac{2\pi}{\hbar}|\langle V'
\rangle|^2\delta(\hbar\omega - E).
$$
The interpretation of this result is that while the photon has definite
momentum and energy, the particle emitting the photon has definite momentum but
indefinite energy. The energy uncertainty comes from the Heisenberg uncertainty
relation $\Delta E \Delta t \ge \hbar$, that is, the shorter the time spent in
state $| 0 \rangle$, the greater the energy uncertainty.

\section{Connection to the Boltzmann Equation in Many-Particle Theory}
\label{sect4}

We have so far been vague about the nature of the interaction $V$. In the
previous section we assumed that we had a set of single-particle states with
a time-independent out-scattering term $V$. In general, however, the particle
of interest will be one of many particles in a system, and $V$ will be an
interaction term between the particles.

The theory of Sections \ref{sect1} and \ref{sect2} is completely general; that
is, the state $| 0 \rangle$ can be taken as a many-particle state instead of a
single-particle state. If it is taken as a many-particle state, however, then
the result (\ref{expg}) is not as useful, since it involves the total
self-energy of the system, while we are typically interested in the
single-particle self-energy.

We define the instantaneous many-particle state using
creation and destruction operators $a_{\bf k}$ and $a^{\dagger}_{\bf k}$ (Baym
\cite{baym}, chapter 19) as
\begin{equation}
| 0 \rangle = \prod_{{\bf k} }
\frac{\left(a^{\dagger}_{\bf k}\right)^{n_{\bf k}}}{\sqrt{n_{\bf k}!}}|
\mbox{vac}
\rangle,
\label{bath}
\end{equation}
where $n_{\bf k}$ gives the instantaneous
occupation number of each state (this is called a ``Fock'' state). The
``vacuum'' state $|\mbox{vac}\rangle$ is the zero-particle state, which in the
case of a solid means the ground state of the system. Note that we do not need
to assume an {\em equilibrium} distribution of particles; we can use an
instantaneous nonequilibrium distribution if we have that information.

A typical interaction term is written in terms of the same creation and
destruction operators, e.g. a two-body, number-conserving term,
\begin{equation}
V = \frac{1}{2}\sum_{{\bf k}_1, {\bf k}_2, {\bf k}_3 } U_{{\bf k}_1, {\bf k}_2,
{\bf k}_3, {\bf k}_4} a^{\dagger}_{{\bf k}_4} a^{\dagger}_{{\bf k}_3}a_{{\bf
k}_2}a_{{\bf k}_1}
\label{int}
\end{equation}
where the summation is not over ${\bf k}_4$ because it is implicitly assumed
that momentum is conserved so that ${\bf k}_4 = {\bf k}_1 + {\bf k}_2 - {\bf
k}_3$. The interaction energy $U$ is assumed to be symmetric on exchange of
${\bf k}_1$ with ${\bf k}_3$ or ${\bf k}_2$ with ${\bf k}_4$.

We are concerned about the evolution of the single-particle state ${\bf k}$,
which is to say, the number of particles in state ${\bf k}$ as a function of
time. The change in the number of particles in a time $t$ is given by
\begin{eqnarray}
d \langle n_{\bf k}\rangle &=& \langle \psi_t | n_{\bf k} | \psi_t \rangle -
\langle 0 | n_{\bf k} | 0\rangle \nonumber\\
&=& \langle \psi(t) |
e^{iH_0t/\hbar} n_{\bf k}e^{-iH_0t/\hbar} | \psi(t)
\rangle - \langle 0 | n_{\bf k} | 0\rangle\nonumber\\
&=& \langle 0 | e^{(i/\hbar)\int V(t) dt}  n_{\bf k} e^{-(i/\hbar)\int V(t)
dt}| 0 \rangle - \langle 0 | n_{\bf k} | 0\rangle
\nonumber\\
&=& \langle 0 | e^{(i/\hbar)\int V(t) dt} {\bf [} n_{\bf k}, e^{-(i/\hbar)\int
V(t) dt} {\bf ]} | 0 \rangle .\label{comm}
\end{eqnarray}
The operator $n_{\bf k}$ commutes with $H_0$, by definition. If it 
commutes with
$V$, then there is no change in $\langle n_{\bf k} \rangle$ over time. We can
resolve the commutator in (\ref{comm}) by using the relations
\begin{eqnarray}
{\bf [} n_{\bf k}, a_{\bf k'} {\bf ]} &=& -a_{\bf k} \delta_{{\bf k},{\bf k'}}
\nonumber
\\ {\bf [} n_{\bf k}, a^{\dagger}_{\bf k'} {\bf ]} &=& a^{\dagger}_{\bf k}
\delta_{{\bf k},{\bf k'}}\label{universal}
\end{eqnarray}
which are valid, surprisingly, for both boson and fermion creation and
destruction operators. For a four-operator term in the interaction (\ref{int}),
we have
\begin{eqnarray}
n_{\bf k}a^{\dagger}_{{\bf k}_4} a^{\dagger}_{{\bf k}_3}a_{{\bf
k}_2}a_{{\bf k}_1} &=&
a^{\dagger}_{{\bf k}_4} a^{\dagger}_{{\bf k}_3}a_{{\bf
  k}_2}a_{{\bf k}_1}\delta_{{\bf k},{\bf k}_4}
+ a^{\dagger}_{{\bf k}_4} a^{\dagger}_{{\bf k}_3}a_{{\bf k}_2}a_{{\bf
k}_1}\delta_{{\bf k},{\bf k}_3}\nonumber\\
&&-a^{\dagger}_{{\bf k}_4} a^{\dagger}_{{\bf k}_3}a_{{\bf k}_2}a_{{\bf
k}_1}\delta_{{\bf k},{\bf k}_2} -a^{\dagger}_{{\bf k}_4} a^{\dagger}_{{\bf
k}_3}a_{{\bf k}_2}a_{{\bf k}_1}\delta_{{\bf k},{\bf k}_1}\nonumber\\
&&+a^{\dagger}_{{\bf k}_4} a^{\dagger}_{{\bf k}_3}a_{{\bf
  k}_2}a_{{\bf k}_1}n_{\bf k}\nonumber
\end{eqnarray}
Thus
\begin{eqnarray}
{\bf [} n_{\bf k}, V {\bf ]} &=& \frac{1}{2}\sum_{{\bf k}_1, {\bf k}_2}
  \left(U_{0321} a^{\dagger}_{{\bf k}}
a^{\dagger}_{{\bf k}_3}a_{{\bf k}_2}a_{{\bf k}_1}
+ U_{3021} a^{\dagger}_{{\bf k_3}}
a^{\dagger}_{{\bf k}}a_{{\bf k}_2}a_{{\bf k}_1}
-U_{3201} a^{\dagger}_{{\bf k}_3}
a^{\dagger}_{{\bf k}_2}a_{{\bf k}}a_{{\bf k}_1}
- U_{3210} a^{\dagger}_{{\bf k}_3}
a^{\dagger}_{{\bf k}_2}a_{{\bf k}_1}a_{{\bf k}} \right)\nonumber\\
&=&\frac{1}{2}\sum_{{\bf k}_1, {\bf k}_2}(U_{D} \pm
U_{E})\left(a^{\dagger}_{{\bf k}} a^{\dagger}_{{\bf k}_1}a_{{\bf k}_2}a_{{\bf
k}_3} - a^{\dagger}_{{\bf k}_3} a^{\dagger}_{{\bf k}_2}a_{{\bf
k}_1}a_{{\bf k}} \right) \label{comm2}
\end{eqnarray}
where $U_{D}$ refers to the direct term and $U_{E}$ to the exchange term, and
the + sign is for bosons and the - sign is for fermions. (For hard-sphere
scattering, i.e. s-wave scattering, $U$ is a constant, which gives a 
factor of 4
enhancement of the scattering cross section for bosons and is forbidden for
fermions.)

We first write out the series expansion,
\begin{eqnarray}
d \langle n_{\bf k}\rangle
  & = & \langle 0 | \left( 1 - (1/i\hbar) \int_{0}^t V(t')dt'
+  ... \right) \nonumber\\
&& \times \left( (1/i\hbar) \int_{0}^t dt'{\bf [ } n_{\bf k} , V(t') {\bf ]}
+   (1/i\hbar)^2 \int_{0}^t dt' \int_{0}^{t'} dt'' {\bf [} n_{\bf k}
,V(t')V(t'') {\bf ]}+ ...
\right) | 0 \rangle
\nonumber
\end{eqnarray}
Any terms in the right-hand series multiplied by the leading ``1'' in the
left-hand series vanish, since $\langle 0 | {\bf [} n_{\bf k}, A
{\bf ]} | 0 \rangle$ vanishes for any operator $A$. Note that
$V^{\dagger}(t) = V(t)$. The leading-order term is therefore
\begin{eqnarray}
d \langle n_{\bf k}\rangle & = & \langle 0 |  (1/\hbar^2) \left(\int_{0}^t
dt' V(t')\right)\left(\int_0^t dt''
  {\bf [ } n_{\bf k} , V(t'') {\bf ]} \right)| 0 \rangle \nonumber\\
&=& \sum_{m }  (1/\hbar^2) \left(\int_{0}^t
dt' e^{(i/\hbar)(E_0-E_m)t'}\right)\left(\int_0^t dt''
e^{(i/\hbar)(E_m-E_0)t''}\right)
\langle 0 | V | m
\rangle \langle m | {\bf [ } n_{\bf k} , V {\bf ]}  | 0 \rangle \nonumber\\
&=& \sum_{m}
\left(\frac{e^{(i/\hbar)(E_0-E_m)t}-1}{E_0-E_m}\right)
\left(\frac{e^{-(i/\hbar)(E_0-E_m)t}-1}{E_0-E_m}\right)
\langle 0 | V | m
\rangle \langle m | {\bf [ } n_{\bf k} , V {\bf ]}  | 0 \rangle \nonumber
\end{eqnarray}
where the sum over states $m$ is over all possible Fock
states. The time-dependent factors are resolved using the identities
\begin{eqnarray}
\lim_{t\rightarrow \infty}\frac{\left( e^{ixt}-1\right) \left(
e^{-ixt}-1\right)}{x^2} &=& \lim_{t\rightarrow \infty}
\frac{\sin^2(xt/2)}{x^2} \nonumber\\
&=& \delta(x)2\pi t
\label{identity}
\nonumber
\end{eqnarray}

We therefore have
\begin{eqnarray}
d \langle n_{\bf k}\rangle  &=&   \sum_m
  \langle 0 | V | m \rangle\langle m |{\bf [} n_{\bf k}, V {\bf
]}| 0 \rangle
\frac{2\pi t}{\hbar}\delta(E_0-E_m).\label{dn}
\end{eqnarray}
Since the different
Fock states are assumed orthonormal, the summation in $V$ is eliminated
because only the terms which couple $|0\rangle$ to $|m\rangle$ survive.
Since a destruction operator $a_{\bf k}$ acting to the right on a state with
$n_{\bf k}$ particles gives a factor $\sqrt{n_{\bf k}}$, and a
creation operator $a^{\dagger}_{\bf k}$ gives a factor $\sqrt{1 \pm n_{\bf
k}}$, where the + sign is for bosons (stimulated emission) and the - sign is
for fermions (Pauli exclusion),  using (\ref{comm2}), (\ref{dn}) therefore
becomes
\begin{eqnarray}
\frac{d \langle n_{\bf k}\rangle}{dt} &=& \frac{2\pi}{\hbar} \sum_{{\bf
k}_1,{\bf k}_2}|U_{D} \pm U_{E}|^2
\left[n_{{\bf k}_3}n_{{\bf k}_2}(1 \pm n_{{\bf k}_1})(1
\pm n_{{\bf k}}) -  n_{{\bf k}}n_{{\bf k}_1}(1 \pm n_{{\bf k}_2})(1 \pm
n_{{\bf k}_3})\right] \nonumber \\
&&\times
\delta(E_{\bf k}+E_{{\bf k}_1} - E_{{\bf k}_3}-E_{{\bf k}_3}) \label{qboltz}
\end{eqnarray}
where we have assumed $t$ is a very small quantity $dt$. This is the quantum
Boltzmann equation, which gives the total rate of
out-scattering for a state  ${\bf k}$.  It is the same energy we would have
written down if we had simply written the total scattering rate as the sum
of the Fermi's Golden Rule rate,
$$
\frac{2\pi}{\hbar}
|\langle m | V | 0 \rangle |^2 \delta(E_0 - E_m),
$$
over all allowed processes of single-particle states $| m \rangle$. It is
correct to order $V^2$, but the limit $t
\rightarrow \infty$  in (\ref{identity}) implies that the time step $dt$
must be long compared to the oscillation time $\hbar/(E_0-E_m)$, i.e.  we are
not concerned about behavior on time scales so short that the equivalent energy
uncertainty is comparable to the typical collision energies. This is the
random-phase approximation again.

We can make one simplification of the Boltzmann equation by treating
$n_{{\bf k}}$ as a continuous variable, in which $n_{{\bf k}}$ is equal to its
average value in an element of phase space $d^3k$, and converting the summation
to an integral, to get
\begin{eqnarray}
\frac{d \langle \overline{n}_{\bf k}\rangle}{dt} &=&
\frac{2\pi}{\hbar}\left(\frac{L^3}{(2
\pi)^3}\right)^2\int d^3k_1 \ d^3k_2    |U_{D} \pm U_{E}|^2
\left[n_{{\bf k}_3}n_{{\bf k}_2}(1 \pm n_{{\bf k}_1})(1
\pm n_{{\bf k}}) -  n_{{\bf k}}n_{{\bf k}_1}(1 \pm n_{{\bf k}_2})(1 \pm
n_{{\bf k}_3})\right]\nonumber\\&&\times
\delta(E_{\bf k}+E_{{\bf k}_1} - E_{{\bf k}_3}-E_{{\bf k}_4}).
\label{avgboltz}
\end{eqnarray}

We can make an additional simplification if we assume that the system is {\em
isotropic}, i.e. the distribution function $n_{{\bf k}}$ depends only on the
magnitude of ${\bf k}$ and not on the direction. Then we can integrate
analytically over all the angles to reduce this integral to just a double
integral which can then be solved numerically for nonequilibrium isotropic
distributions. This has been used to produce predictions for various
nonequilibrium systems \cite{snoke1,snoke2,snoke3,snoke4}.

These different rates clarify the difference between $T_1$, $T_2$ etc. in
experiments. The rate (\ref{qboltz}) gives the rate of depletion of a single
quantum state, sometimes called the ``dephasing'' rate in
optics. The time constant for this decay is called ``$T_2$'' in NMR
terminology. The rate (\ref{avgboltz}) gives the rate of depletion of 
states with
the same macroscopic characteristics as state
${\bf k}$, and the time constant is called ``$T_1$'' in NMR terminology.

We now have calculated the many-body equivalent of the formula (\ref{expg}) for
the decay rate. We can now use this in the calculation of line broadening.

\section{Line broadening in Many-Particle theory}

As in Section \ref{sect3}, we are interested in the rate of emission of a
photon with momentum ${\bf k}$ and frequency $\omega$ from a given initial
state, via a weak particle-photon interaction in the presence of a
particle-particle interaction. In the many-particle theory we will define the
creation and destruction operators $c^{\dagger}_{\bf k}$ and
$c_{\bf k}$ for photons; the total many-particle state is a product of Fock
states of both photons and the interacting particles. As before, we write
$H=H_0+V+V'$, where
$V'$ is the particle-photon interaction term, which we assume has the form
$$
V' = \sum_{{\bf k}_1} P_{{\bf k}_1} c^{\dagger}_{{\bf k}_1}a_{{\bf k}_1}.
$$
Although this form is strictly unphysical, it is a good model for radiative
transitions. Electrons are, of course, conserved, but if we restrict 
our attention to
only one band, then radiative transitions which cause band-to-band transitions
involve the destruction of an electron in one band. On the other 
hand, excitons are
not conserved, so this notation allows us to treat electrons and 
excitons on equal
footing as in the previous section. Also, we drop the hermitian conjugate term
because we neglect the possibility of photon absorption; we assume there are no
photons at $t=0$.

The number of photons emitted in a time $t$ is then
\begin{eqnarray}
\langle N_{\bf k}\rangle & =&  \langle  \psi(t) |
c^{\dagger}_{\bf k}c_{\bf k}  |\psi(t) \rangle
\nonumber\\ &=& \langle  0 | e^{(i/\hbar)\int (V(t)+V'^{\dagger}(t)) 
dt} {\bf [}
c^{\dagger}_{\bf k}c_{\bf k},e^{-(i/\hbar)\int (V(t)+V'(t)) dt} {\bf ]} |0
\rangle
\end{eqnarray}
As in the previous section, we assume $|0\rangle$ is an exact Fock 
state which is
not an eigenstate of the Hamiltonian. This will complicate the 
analysis (preventing
us from using standard many-body formalism) but allows us to treat 
the case of a
highly nonequilibrium system. As before, we write out the series,
\begin{eqnarray}
\langle N_{\bf k}\rangle & = &
\langle 0 |
\left( 1 - (1/i\hbar) \int_{0}^t
\left(V(t')+V'^{\dagger}(t')\right)dt' \right. \nonumber\\
&& \left. +  (1/i\hbar)^2
\int_{0}^t dt'
\int_{0}^{t'} dt''  \left(V(t'')+V'^{\dagger}(t'')
\right)\left(V(t')+V'^{\dagger}(t')
\right)...
\right)
\nonumber\\ && \times \left((1/i\hbar) \int_{0}^t dt'
  {\bf [}c^{\dagger}_{\bf k}c_{\bf k} ,(V(t')+V'(t')){\bf ]} \right.
\nonumber\\ && \left. +   (1/i\hbar)^2 \int_{0}^t dt'
\int_{0}^{t'} dt''
{\bf [}c^{\dagger}_{\bf k}c_{\bf k}, \left(V(t')+V'(t')
\right)\left(V(t'')+V'(t'') \right){\bf ]} + ...
\right) | 0 \rangle
\nonumber
\end{eqnarray}
Since $V'$ is not number-conserving, only terms with products of the
pair $V'^{\dagger}V'$ survive. We restrict ourselves to only terms which are
first order in $V'^{\dagger}V'$. The photon number
$c^{\dagger}_{\bf k}c_{\bf k}$ commutes with
$V(t)$, while the relation (\ref{universal}) implies that
$[c^{\dagger}_{\bf k}c_{\bf k},V'] =  P_{{\bf k}}c^{\dagger}_{{\bf k}}a_{\bf
k}$,  which in turn implies that only terms with
$V'^{\dagger}=P^*_{{\bf k}}a^{\dagger}_{\bf k}c_{{\bf k}}$ survive.

At this point we can make an additional assumption in order to 
greatly simplify the
calculation. This is to assume the ``dilute'' limit, which is that 
the occupation
number of state ${\bf k}$ in $|0\rangle$ is not greater than one, and 
in-scattering
to any particular state ${\bf k}$ is negligible compared to 
out-scattering. One can
see that this will be the case because for any out-scattering term 
$a^{\dagger}_{{\bf
k}} a^{\dagger}_{{\bf k}_3} a_{{\bf k}_2} a_{{\bf k_1}}$ for which 
the four momenta
satisfy momentum conservation exactly, the probability of having both 
of states ${\bf
k}_1$ and ${\bf k}_2$ occupied simultaneously is very low.  Note that in
equilibrium, the average net scattering into and out of any state over a long
time equals zero, but in this case the many-body state $| 0\rangle$ 
has been prepared
with $n_{\bf k} = 1$ at $t=0$, and therefore the out-scattering {\em 
at that time}
will greatly exceed the in-scattering.

In terms like $VV'$ there are combinations like
$a^{\dagger}_{{\bf k}_4} a^{\dagger}_{{\bf k}_3} a_{{\bf k}_2} 
a_{{\bf k}_1}a_{\bf
k}$. No double occupancy means that terms with $a_{{\bf k}}a_{{\bf 
k}}$ vanish.  On
the other hand,  terms in $V$ for which no ${\bf k}_i = {\bf k}$ commute with
$a_{\bf k}$ and one can show that all such terms vanish.  We will use 
$V_{\bf k}(t)$
to indicate the subset of terms in the summation of $V(t)$ which do not commute
with
$a_{\bf k}$. We then get
\begin{eqnarray}
\langle N_{\bf k}\rangle & = &
\langle 0 | |P_{\bf k}|^2
\left(  - (1/i\hbar) \int_{0}^t a^{\dagger}_{\bf
k}c_{{\bf k}}e^{-(i/\hbar)(\hbar\omega-E_{\bf k})t'} dt'
+(1/i\hbar)^2 \int_{0}^t dt' \int_{0}^{t'} dt''
V_{\bf k}(t'') a^{\dagger}_{\bf
k}c_{{\bf k}}e^{-(i/\hbar)(\hbar\omega-E_{\bf k})t'}
\right.\nonumber\\ && \left. -  (1/i\hbar)^3 \int_{0}^t dt' \int_{0}^{t'}
dt''\int_{0}^{t''} dt''' V_{\bf k}(t''')V_{\bf k}(t'')a^{\dagger}_{\bf
k}c_{{\bf k}}e^{-(i/\hbar)(\hbar\omega-E_{\bf k})t'} +-...
\right)
\nonumber\\ && \times \left( (1/i\hbar) \int_{0}^t dt'
  c^{\dagger}_{{\bf k}}a_{\bf k}e^{(i/\hbar)(\hbar\omega-E_{\bf
k})t'}
  +   (1/i\hbar)^2 \int_{0}^t dt'
\int_{0}^{t'} dt''
c^{\dagger}_{{\bf k}}a_{\bf k}e^{(i/\hbar)(\hbar\omega-E_{\bf k})t'}
V_{\bf k}(t'')
\right. \nonumber\\ &&
\left. + (1/i\hbar)^3 \int_{0}^t dt' \int_{0}^{t'} dt''\int_{0}^{t''} dt'''
c^{\dagger}_{{\bf k}}a_{\bf k}e^{(i/\hbar)(\hbar\omega-E_{\bf
k})t'} V_{\bf k}(t'')V_{\bf k}(t''') +...
\right) | 0 \rangle
\nonumber\\
&=& |P_{\bf k}|^2 \langle 0|
\left(
-(1/i\hbar) \int_{0}^t e^{-(i/\hbar)(\hbar\omega-E_{\bf k})t'}
dt'  \right)\nonumber\\&&
\times\left(1-(1/i\hbar)\int_0^{t'} V_{\bf k}(t'') dt'' + (1/i\hbar)^2
\int_0^{t'}\int_0^{t''}V_{\bf k}(t'')V_{\bf k}(t''') dt''dt'''
-+...\right)\nonumber\\&&
\times
n_{\bf k}\left(
(1/i\hbar) \int_{0}^t e^{(1/i\hbar)(\hbar\omega-E_{\bf k})t'}dt'\right)
\nonumber\\&&
\times  \left(1+(1/i\hbar)\int_0^{t'} V_{\bf k}(t'') dt'' + (1/i\hbar)^2
\int_0^{t'}\int_0^{t''}V_{\bf k}(t'')V_{\bf k}(t''') dt''dt''' +...\right)
|0\rangle
\nonumber\\
&=&
\frac{|P_{\bf k}|^2 }{\hbar^2}\langle 0|
\left(
  \int_{0}^t e^{-(i/\hbar)(\hbar\omega-E_{\bf k})t'}
dt'  \right)e^{(i/\hbar)\int_0^{t'} V_{\bf k}(t'') dt''}
\nonumber\\&& \times  n_{\bf k}
\left(
  \int_{0}^t e^{(1/i\hbar)(\hbar\omega-E_{\bf k})t'}dt'\right)
  e^{-(i/\hbar)\int_0^{t'} V_{\bf k}(t'') dt''}
|0\rangle
\end{eqnarray}
We now break this into two parts, inserting a complete set of states
$\sum|m\rangle\langle m|$ and separating the term with $|0\rangle\langle 0 |$
from the rest.
\begin{eqnarray}
\langle N_{\bf k}\rangle&=&
\frac{|P_{\bf k}|^2 }{\hbar^2}\Biggl[ \left(
  \int_{0}^t e^{-(i/\hbar)(\hbar\omega-E_{\bf k})t'}
dt'  \langle 0 | e^{(i/\hbar)\int_0^{t'} V_{\bf k}(t'') dt''} |0\rangle\right)
\nonumber\\&&\times n_{\bf k} \left(
  \int_{0}^t e^{(1/i\hbar)(\hbar\omega-E_{\bf k})t'}dt'
\langle 0 |   e^{-(i/\hbar)\int_0^{t'} V_{\bf k}(t'') dt''}
|0\rangle\right)
  \nonumber\\&&\left.
+\sum_{m \ne 0}  (1/\hbar^2)\left| \int_0^t
dt'e^{-(i/\hbar)(\hbar\omega-E_{\bf k})t'}
\int_0^{t'}dt''  \langle 0 |V_{\bf k}(t'') |m
\rangle\right|^2+... \right]
\end{eqnarray}
The second, cross term vanishes, having terms like
$$
\left| \int_0^t dt' e^{i\omega_1 t'}\frac{e^{i\omega_2 
t'}-1}{\omega_2} \right|^2
= \left|
\frac{e^{i(\omega_1+\omega_2)t}-1}{(\omega_1+\omega_2)\omega_2}-\frac{e^{i\omega_1
t}}{\omega_1\omega_2}\right|^2
$$
which vanish
  in the limit of the random-phase approximation,
while the other factors are given by (\ref{exp}) -- (\ref{expg}).  In the
limit $t\rightarrow \infty$, we then have
\begin{eqnarray}
\langle N_{\bf k}\rangle &=&
\frac{|P_{\bf k}|^2 }{\hbar^2}\langle n_{\bf k}\rangle \left|
  \int_{0}^{\infty} dt' e^{(i/\hbar)(\hbar\omega-E_{\bf k})t'}
    e^{-(i/\hbar)(\Delta^{(1)}+\Delta^{(2)} -i\Gamma^{(2)}  )} \right|^2
\nonumber\\ &=& |P_{\bf k}|^2 \langle n_{\bf k} \rangle \frac{1}{(\hbar\omega -
E)^2+\Gamma^2},
\label{resolution}
\end{eqnarray}
where we have used the abbreviated notation for $E$ and $\Gamma$  of Section
\ref{sect3}, and
\begin{eqnarray}
\Gamma^{(2)} &=& \pi\sum_{m\ne 0}|\langle m | V_{\bf k} | 0
\rangle|^2 \delta(E_0-E_m) \nonumber\\
&=& \pi\sum_{m\ne 0}\left|\langle m | \frac{1}{2}\sum_{{\bf k}_1,{\bf 
k}_2} (U_D \pm
U_E)
a^{\dagger}_{{\bf k}_3}a^{\dagger}_{{\bf k}_2}a_{{\bf k}_1}a_{\bf k} | 0
\rangle\right|^2 \delta(E_0-E_m) \nonumber\\
&=&  \pi \sum_{{\bf
k}_1,{\bf k}_2}|U_{D} \pm U_{E}|^2
   n_{{\bf k}}n_{{\bf k}_1}(1 \pm n_{{\bf k}_2})(1 \pm
n_{{\bf k}_3})
\delta(E_{\bf k}+E_{{\bf k}_1} - E_{{\bf k}_3}-E_{{\bf k}_3})
\label{self-energy}
\end{eqnarray}
Note that the integral in (\ref{self-energy}) is the same as in (\ref{qboltz}),
neglecting the in-flow  terms, which are negligible in the dilute limit for a
{\em particular} state ${\bf k}$, even though the average in-flow to 
a region of
phase space $d^3k$ around state ${\bf k}$, found in (\ref{avgboltz}), 
equals the
out-flow from the same region in equilibrium.

We can rewrite (\ref{resolution}) as
\begin{eqnarray}
\frac{d\langle N_{\bf k}\rangle}{dt} &=&
  |P_{\bf k}|^2 \frac{d \langle n_{\bf k}\rangle}{dt} \frac{1}{(\hbar\omega -
E)^2+\Gamma^2} \nonumber\\
&=& |P_{\bf k}|^2  \frac{2\Gamma/\hbar}{(\hbar\omega -
E)^2+\Gamma^2}
\label{newbroad}
\end{eqnarray}
in which (\ref{qboltz}) is used to obtain $d\langle n_{\bf k}\rangle/dt \equiv
-2\Gamma/\hbar$.

This all has been essentially a long justification of a very simple formula,
which we already obtained in (\ref{rate}) using the single-particle picture.
This calculation shows us the limits of validity of this formula,
however. In particular, it assumes the dilute limit. In the gas of a
high-density boson gas, the situation is much more complex, as shown by
Shi, Verachaka and Griffin \cite{griffin} for case of a 
quasiequilibrium boson gas.
When state
${\bf k}$ is nondegenerate, however, this calculation shows that we can use the
the interaction term
$V' =
\sum |U_D \pm U_E| \sqrt{n_1}\sqrt{1\pm n_2}\sqrt{1\pm n_3}$ with 
generality in the
single-particle picture for a particle in interacting with a bath that is in
equilibrium or far from equilibrium. This type of interaction 
potential has been used
to predict the Lorentzian line broadening of luminescence from 
various semiconductors
\cite{lauten,snoke-shields}, known as ``homogeneous'' broadening (as opposed to
``inhomogeneous'' broadening, due to random fluctuations of the band gap.)

An even simpler method, which is justified by this calculation, is to simply
write $H=H_0' + V'$ where $H_0'$ includes the self-energy correction to the
single-particle energy due to scattering by
$V$. In this case we would write $e^{(i/\hbar)H_0't}| 0 \rangle =
e^{(i/\hbar)(E_{\bf k}+\Delta_{\bf k}^{(1)}+
\Delta_{\bf k}^{(2)}- i\Gamma^{(2)})t}$ and then treat the interaction $V'$
in first-order perturbation theory.

An important implication of the calculations here is that even in the case when
there is substantial line broadening, the energy conserving delta-function in
(\ref{qboltz}) should {\em not} be replaced by a Lorentzian. One can 
show that if
one were to do this, total energy conservation of the system would be violated.

\section{The Connection to $\chi^{(3)}$ in Four-Wave Mixing}

We have seen that the imaginary self energy gives the line broadening of
optical transitions. Alternatively, as seen above, it also corresponds to the
out-scattering rate. This rate can often be measured directly in time-resolved
measurements. The two methods are complementary. If the rate is extremely fast,
it may be too fast for time-resolved methods, but then it gives signficant line
broadening. On the other hand, if the line broadening is too small for the
experimental spectral resolution to pick up, it corresponds to slow decay rate
in time-resolved measurements.

The nonlinear susciptibility is defined in terms of Maxwell's wave equation in
the presence of a polarization current $J$,
\begin{equation}
\frac{\partial^2 E}{\partial x^2} = \mu_0\epsilon_0 \frac{\partial^2 
E}{\partial t^2} +
\mu_0 \frac{\partial J}{\partial t}.
\end{equation}
The polarization current depends on the electric field in a complex 
way, which is
represented to third order as
\begin{equation}
J  =  \epsilon_0 \frac{\partial }{\partial
t} \left( \chi E + \chi^{(2)}E^2 + \chi^{(3)}E^3 \right).
\end{equation}
To calculate $\chi^{(3)}$ from first principles, then, we need to 
calculate the
current due to the oscillating dipole moment of the medium. If we 
recall that the
{\bf J} = e{\bf V}  = e{\bf P}/m, we see that fundamentally we want 
to calculate the
momentum $\langle P\rangle$ of the oscillator as a function of electric field.
The value of $\chi^{(3)}$ can then be identified as the part of this 
which depends to
the third order on the electric field $E$.

For simplicity we will assume that the oscillator has two states which
couple only weakly to other states.  These two states nominally
correspond to the valence and conduction electron bands in a semiconductor. We
will label the ground state $|v\rangle$ and the excited state $|c\rangle$.  We
will also assume that the light field couples only the valence band to the
excited band and has no effect on coupling within bands.

Having justified its use in the previous section, we use the simple approach of
replacing each state's energy with the appropriate renormalized 
energy including
imaginary self-energy due to intraband interactions, subject to the same
constraint of no multiple occupancy of states, and treat the optical 
transitions in
first-order perturbation theory.
  Let the time-dependent electron state be $|e_t\rangle$. In the
interaction representation we use $|e(t)\rangle = e^{iH_0t/\hbar}|e_t\rangle$.
Then the oscillator momentum of interest is
\begin{eqnarray}
P(t) &=& \langle e_t|P| e_t \rangle = \langle
e(t)|e^{iH_0t/\hbar}Pe^{-iH_0t/\hbar}|e(t)\rangle \\
\nonumber
&=&\langle  e(t) |e^{iH_0t/\hbar} \sum_i |i\rangle\langle i| P \sum_j
|j\rangle\langle j|e^{-iH_0t/\hbar} |e(t)\rangle\\
\nonumber &=& \langle e(t) |c\rangle e^{i(E_c+i\Gamma_c)t/\hbar} \langle
c|P|v\rangle
\langle v|e(t)\rangle e^{-i(E_v-i\Gamma_v)t/\hbar} \\
\nonumber
&&+\langle e(t)|v\rangle e^{i(E_v+i\Gamma_v)t/\hbar}\langle v|P|c\rangle\langle
c|e(t)\rangle e^{-i(E_c-i\Gamma_c)t/\hbar}\\  &=& 2\Re \langle v|P|c\rangle
\langle e(t)|v\rangle\langle c|e(t)\rangle e^{i(E_v-E_c)t/\hbar}e^{-(\Gamma_v +
\Gamma_c)t/\hbar},
\nonumber
\end{eqnarray}
where $\langle v|P|c\rangle$ is the standard oscillator strength of the
transition.

The time dependence of the electronic state in response to the 
electric field is
determined by the  Schr\"odinger equation
\begin{equation}
i\hbar\frac{\partial }{\partial t}| e_t\rangle = H | e_t \rangle
\end{equation}
\begin{eqnarray}
H &=& H_0 + V' \\
V' &=& \frac{e}{2mc} P\cdot A\\
E &=& -\frac{1}{c}\frac{\partial A}{\partial t}.
\end{eqnarray}
$V'$ is the radiation term which connects the
valence and conduction band states, and $H_0$ is the total Hamiltonian of
the electron states. $A$ and $E$ are assumed to be classical fields. 
Let $A(t) = A_0
e^{-i\omega t}$. Then
\begin{eqnarray}
E(t)  &=& (i/c) A_0 \omega e^{-i \omega t} =
E_0 e^{-i\omega t}
\end{eqnarray}
  which implies
\begin{equation}
V' = \frac{e P\cdot E_0}{2m\omega}e^{-i\omega t}
\end{equation}
In the interaction representation,
\begin{equation}
V'(t) = e^{iH_0t/\hbar}\frac{e P E_0}{2m\omega}e^{-i\omega t} e^{-iH_0t/\hbar}.
\end{equation}
Third-order time-dependent perturbation theory then gives
\begin{eqnarray}
|e(t) \rangle & =& | e(t_0) \rangle  + \frac{1}{i\hbar} \int^t_{t_0} 
dt' V'(t') |e(t_0)\rangle  \\
\nonumber &&+ \frac{1}{(i\hbar)^2} \int^t_{t_0} dt' \int^{t'}_{t_0} dt''
V'(t')V'(t'') |e(t_0)\rangle  \\ \nonumber
&&+ \frac{1}{(i\hbar)^3} \int^t_{t_0} dt' \int^{t'}_{t_0} dt'' \int^{t''}_{t_0}
dt''' V'(t')V'(t'')V'(t''') |e(t_0)\rangle
\end{eqnarray}
where $t_0 \rightarrow -\infty$.

Let $|e(t_0)\rangle  = | v\rangle $. Then since we assume that $V'$ does
not couple intraband states, so that only
$\langle c|V'|v\rangle $ terms survive, we have
\begin{eqnarray}
\langle c|e(t) \rangle & =&  \frac{1}{i\hbar} \int^t_{t_0} dt' \langle c|V'(t')
|v\rangle  \\ \nonumber &&+ \frac{1}{(i\hbar)^3} \int^t_{t_0} dt'
\int^{t'}_{t_0} dt'' \int^{t''}_{t_0} dt''' \langle c|V'(t')|v\rangle \langle
v|V'(t'')|c\rangle \langle c|V'(t''') |v\rangle  \\ \nonumber
\langle v|e(t) \rangle & =& 1
+ \frac{1}{(i\hbar)^2} \int^t_{t_0} dt' \int^{t'}_{t_0} dt''
\langle v|V'(t')|c\rangle \langle c|V'(t'') |v\rangle
\end{eqnarray}
Therefore
\begin{eqnarray}
P(t) &=& 2 \Re\langle v|P|c\rangle e^{i(E_v-E_c)t/\hbar}e^{-(\Gamma_v +
\Gamma_c)t/\hbar} \left[\frac{1}{i\hbar}
\int^t_{t_0} dt' \langle c|V'(t') |v\rangle
\right. \\
\nonumber &&+ \frac{1}{(i\hbar)^3} \int^t_{t_0} dt' \int^{t'}_{t_0} dt''
\int^{t''}_{t_0} dt''' \langle c|V'(t')|v\rangle \langle 
v|V'(t'')|c\rangle \langle c|V'(t''') |v\rangle  \\ \nonumber
&&+ \frac{1}{(i\hbar)^3} \left( \int^t_{t_0} dt' \langle c|V'(t') 
|v\rangle  \right)
\left( \int^t_{t_0} dt' \int^{t'}_{t_0} dt''
\langle c|V'(t')|v\rangle \langle v|V'(t'') |c\rangle  \right) \\ \nonumber
&&+ \biggl.  {\cal O}(V'^5) \biggr]
\end{eqnarray}
There is no second-order $\chi^{(2)}$ by symmetry. We will concentrate on the
first third-order term; the calculation of the second term gives 
similar results.

Substituting in for $V'(t)$, and performing the time integrals 
explicitly, we get
\begin{eqnarray}
P(t) &=& 2 \Re \left[  |\langle v|P|c\rangle |^2
e^{i(E_v-E_c)t/\hbar}e^{-(\Gamma_v +
\Gamma_c)t/\hbar}\frac{eE_0}{2m\omega}  \frac{1}{i\hbar}
\int^t_{t_0} dt' e^{i(E_c-i\Gamma_c)t'/\hbar}e^{- i\omega
t'}e^{-i(E_v+i\Gamma_v)t'/\hbar}
\right. \nonumber \\
\nonumber &&+ |\langle v|P|c\rangle |^4
e^{i(E_v-E_c)t/\hbar}e^{-(\Gamma_v +
\Gamma_c)t/\hbar}\left(\frac{eE_0}{2m\omega}\right)^3
\frac{1}{(i\hbar)^3}
\int^t_{t_0} dt' \int^{t'}_{t_0} dt'' \int^{t''}_{t_0} dt''' \\ \nonumber
&&\times \left(e^{i(E_c-i\Gamma_c)t'/\hbar}e^{- i\omega
t'}e^{-i(E_v+i\Gamma_v)t'/\hbar} \right) \left(
  e^{i(E_v-i\Gamma_v)t''/\hbar}e^{- i\omega 
t''}e^{-i(E_c+i\Gamma_c)t''/\hbar}\right)
\\ \biggl.
&&\times \left(e^{i(E_c-i\Gamma_c)t'''/\hbar}e^{- i\omega
t'''}e^{-i(E_v+i\Gamma_v)t'''/\hbar}\right)
\biggr]
\label{ptint} \\ \nonumber
&=&  2 \Re\biggl[ |\langle v|P|c\rangle |^2
e^{i(E_v-E_c)t/\hbar}e^{-(\Gamma_v +
\Gamma_c)t/\hbar}\frac{eE_0}{2m\omega}
\frac{e^{i(E_c- \hbar\omega -E_v)t/\hbar}}{(E_c - E_v -\hbar\omega +
i(\Gamma_c+\Gamma_v))} e^{(\Gamma_v +
\Gamma_c)t/\hbar} \biggr.\\ \nonumber
&&+ |\langle v|P|c\rangle |^4
e^{i(E_v-E_c)t/\hbar}e^{-(\Gamma_v +
\Gamma_c)t/\hbar}\left(\frac{eE_0}{2m\omega}\right)^3
\frac{1}{(i\hbar)^2}
\int^t_{t_0} dt' \int^{t'}_{t_0} dt''  \\ \nonumber
&&\times \left( e^{i(E_c-i\Gamma_c)t'/\hbar}e^{-i\omega
t'}e^{-i(E_v+i\Gamma_v)t'/\hbar} \right) \left(
  e^{i(E_v-i\Gamma_v)t''/\hbar}e^{- i\omega 
t''}e^{-i(E_c+i\Gamma_c)t''/\hbar} \right)
\\  \biggl.
&&\times
\frac{e^{i(E_c- \hbar\omega -E_v)t''/\hbar}}{(E_c - E_v
-\hbar\omega + i(\Gamma_c+\Gamma_v))}e^{(\Gamma_v +
\Gamma_c)t''/\hbar}\biggr]\\ \nonumber
&=& 2 \Re\biggl[ |\langle v|P|c\rangle |^2
\frac{eE_0e^{-i \omega t}}{2m\omega}
\frac{1}{(E_c - E_v -\hbar\omega +
i(\Gamma_c+\Gamma_v))}  \biggr.\\ \nonumber
&&+ |\langle v|P|c\rangle |^4
\left(\frac{eE_0e^{-i\omega t}}{2m\omega}\right)^3
\frac{1}{(E_c - E_v
-3\hbar\omega + i3(\Gamma_c+\Gamma_v))}
\left(\frac{ 1}{2\omega}\right) \\
&& \times \biggl.
\frac{1}{(E_c - E_v
-\hbar\omega + i(\Gamma_c+\Gamma_v))}e^{2(\Gamma_c+\Gamma_v)t/\hbar}\biggr]
\label{pt}
\end{eqnarray}
The first term of the final result is the linear $\chi$ which goes 
into the index of
refraction, and the second term is proportional to $E^3$, i.e. it 
gives $\chi^{(3)}$.
Here we have deduced only a frequency-tripling nonlinear effect with 
an ingoing and
outgoing resonance, because we assumed only one input frequency. Of 
course, if we
had  written $A(t) =
A_1(e^{i\omega_1t}+e^{-i\omega_1t})
+A_2(e^{i\omega_2t}+e^{-i\omega_2t})+A_3(e^{i\omega_3t}+e^{-i\omega_3t})$ 
instead of
$A(t) = A_0e^{-i\omega t}$ as we did, then we would get all the 
frequency mixing
terms associated with third-order optics, e.g. four-wave mixing (FWM).

As seen in (\ref{ptint}), the prefactor
$e^{i(E_v-E_c)t/\hbar}e^{-(\Gamma_v +\Gamma_c)t/\hbar}$ in each of the terms is
exactly canceled out by a term with opposite sign in the exponent from the
time integrals, so that the oscillation of the polarization is only 
at the driving
frequency and the third-order harmonics.

Suppose that instead of continuing on forever,
the electric field is shut off at $t=0$, i.e.
\begin{equation}
A(t) = \left\{ \begin{array}{cc}
A_0e^{-i\omega t}, &  t<0\\
0, & t> 0
\end{array} \right.
\end{equation}
Then for times $t>0$, the integrals in (\ref{ptint}) are time-independent. The
only time dependence in $P(t)$ comes from the 
$e^{i(E_v-E_c)t/\hbar}e^{-(\Gamma_v +
\Gamma_c)t/\hbar}$ prefactor.
The polarization continues to oscillate at frequency 
$(E_v-E_c)/\hbar$, which is
resonant at both $\omega$ and $3\omega$ (the sum frequency). If there 
is no scattering
(damping) it will continue ringing forever. The imaginary part of the 
self energy
gives the rate at which the phase-coherent polarization dies. This is 
why it can be
called a ``dephasing'' rate. There is also the feature, seen in 
(\ref{pt}), that the
polarization grows exponentially in time at twice the decay rate, 
which implies the
counterintuitive prediction that a FWM signal will have a longer rise time if
there is less dephasing.  This has been seen experimentally
\cite{buildup1,buildup2}.

In the case of delayed FWM, a third, ``probe'' wave follows
two ``pump'' waves by some time delay. The strength of the fourth 
wave then gives a
measure of the $T_2$ time. This is because the first two waves each create a
polarization wave in the medium due to the {\em linear} term in 
(\ref{ptint}), which
decays according to the prefactor $e^{-(\Gamma_v +\Gamma_c)t/\hbar}$. 
These two waves
then become contributions to $E$ at later times, which mix with the 
probe wave via the
the third-order term. Since the strengths of the pump polarization 
waves decay, so
will the delayed FWM signal, with exactly the same time constant.

We have seen then that the FWM dephasing rate and the line broadening are both
controlled by the same imaginary self-energy, i.e. the $T_2$ time. 
This leads one
to expect that the line width and the FWM dephasing rate should be inversely
proportional. This is generally true, but can be incorrect when a continuum of
states is excited, (e.g. by an ultrafast pulse with considerable Heisenberg
energy uncertainty), which leads to interference between different oscillator
frequencies, as seen in Ref. \cite{wegener}.

\section{Scaling Laws in Recent Experiments}

The electron-electron Coulomb scattering process represents a special case of
the two-body scattering discussed in Section \ref{sect4}, because the
interaction cross section diverges in the case of zero momentum transfer, and
this divergence must be removed self-consistently by a screening length which
depends on the density and the instantaneous energy distribution of the
electrons. A previous publication
\cite{snoke4}  presented the results of scaling laws  for electron-electron
scattering based on a Boltzmann integral calculation taking into account the
dependence of the self-consistent screening length on the density, using a
method valid both for nonequilibrium and equilibrium electron plasmas. The
main conclusions of that work are reviewed here.

First, the proper integral for the electron-electron scattering 
depends on the type of
experiment. There are three different integrals which relate to the 
experiments. The
first is the ``total'' scattering rate, i.e. the imaginary 
self-energy calculated in
(\ref{self-energy}) above. This integral determines the rate of decay of
$\chi^{(3)}$, as shown above. A second integral is given by (\ref{avgboltz})
weighted by the momentum exchanged by the electrons in a given scattering
event. This typically controls the electron scattering rate determined in
transport measurements. A third integral is given by (\ref{avgboltz}) weighted
by the energy exchanged by the electrons in a given scattering event. This
determines the evolution of the carrier distribution function, which is
typically recorded in time-resolved luminescence experiments. These 
three rates are
approximately the same for hard-sphere scattering but are very different for
Coulomb scattering. This means that the ``relaxation time 
approximation,'' in which
all scattering processes are assumed to be characterized by a single 
``relaxation
time,'' is justified in the case of short-range interactions but breaks down
completely for long-range Coulomb scattering.

\begin{table}
\begin{tabular}{lcc}
& dephasing  ($T_2$)      & energy  relaxation ($T_1$)     \\ \hline
low density: \\
3d & constant \cite{arlt} & $n^{1/2}$ \cite{snoke1,hannak,kash1}\\
2d & constant  & $n$ \cite{kash2}  \\
\\
high density: \\
3d &  $n^{1/3}$ \cite{shank1} & $n^{2/3}$   \\
2d &  $n^{1/2}$ \cite{shank2} & $n$ \cite{kash2}  \\
\end{tabular}
\caption{Density dependence of rates for electron-electron scattering 
predicted in
Ref. \protect\cite{snoke4}. Experimental confirmations are given in the
references in the table.}
\end{table}

Second, the scaling laws as a function of temperature and density for 
these various
integrals were determined for both two dimensions and three 
dimensions. At that time,
several experiments had been done which gave scaling laws consistent 
with the results
of the calculations; recent experiments also fulfill the
predictions of that theory. Table 1 gives a summary of the predicted density
dependences and the experiments which have observed them. An 
important result is that
this theory predicts that the dephasing rate is independent of 
density at low density,
in basic agreement with recent experiments (Ref. \cite{arlt}; also 
seen indirectly in
Ref. \cite{wegener2}.)  The ``high-density regime'' is defined as the 
regime in which
the classical screening length becomes comparable to the 
interparticle spacing. A full
treatment of the scattering rate in this regime would require 
accounting for the
quantum wavefunctions; however, the proper scaling law for the 
dephasing can be found
by the simple assumption that the screening length is pinned at the 
interparticle
spacing. This assumption also correctly gives the crossover from the 
high-density to
low-density scaling regimes.

A third main conclusion of Ref. \cite{snoke4} was that the screening length of
the electrons scales with density in same way even for a highly nonequilibrium
distribution. Quantum memory effects are important in determining the 
exact evolution
of polarization, but the scaling laws should remain the same even 
when quantum memory
effects are taken into account.

Recent experiments showing a near-constant dephasing rate at low 
density have been
interpreted in terms of optical phonon scattering \cite{arlt}. A full study of
dephasing as a function of the excition photon energy and the 
temperature should
distinguish between these two interpretations, since Ref. 
\cite{snoke4} also gave
predictions for the temperature dependence of the scaling laws.

\section{Conclusions}

The iostropic quantum Boltzmann equation has been used to produce quantitative
predictions for numerous experimental systems far from equilibrium
\cite{snoke1,snoke2,snoke3,snoke4}. These calculations show that the integrals
which enter into these calculations are the same as the integrals 
which are used for
optical line broadening and four-wave mixing, for the cases of a Fermi or
nondegenerate Bose gas. In the case of electron Coulomb scattering, 
great care must
be taken to use the properly weighted integral for different experiments.

{\bf Acknowledgements.}  This work has been supported by the National Science
Foundation as part of Early Career award DMR-97-22239. One of the 
authors (D.S.) is a
Cottrell Scholar of the Research Corporation.

\end{document}